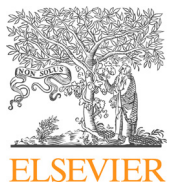
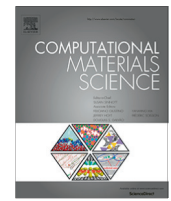

# Migration barriers for surface diffusion on a rigid lattice: Challenges and solutions

Ekaterina Baibuz [a,*], Simon Vigonski [b,a], Jyri Lahtinen [a], Junlei Zhao [a], Ville Jansson [a], Vahur Zadin [b,a], Flyura Djurabekova [a,c]

[a] Helsinki Institute of Physics and Department of Physics, P.O. Box 43 (Pietari Kalmin Katu 2), FI-00014 University of Helsinki, Finland
[b] Institute of Technology, University of Tartu, Nooruse 1, 50411 Tartu, Estonia
[c] National Research Nuclear University MEPhI, Kashirskoye sh. 31, 115409 Moscow, Russia

## ABSTRACT

Atomistic rigid lattice Kinetic Monte Carlo is an efficient method for simulating nano-objects and surfaces at timescales much longer than those accessible by molecular dynamics. A laborious part of constructing any Kinetic Monte Carlo model is, however, to calculate all migration barriers that are needed to give the probabilities for any atom jump event to occur in the simulations. One of the common methods of barrier calculations is Nudged Elastic Band. The number of barriers needed to fully describe simulated systems is typically between hundreds of thousands and millions. Calculations of such a large number of barriers of various processes is far from trivial. In this paper, we will discuss the challenges arising during barriers calculations on a surface and present a systematic and reliable tethering force approach to construct a rigid lattice barrier parameterization of face-centred and body-centred cubic metal lattices. We have produced several different barrier sets for Cu and for Fe that can be used for KMC simulations of processes on arbitrarily rough surfaces. The sets are published as Data in Brief articles and available for the use.

© 2018 The Authors. Published by Elsevier B.V. This is an open access article under the CC BY license (http://creativecommons.org/licenses/by/4.0/).

## 1. Introduction

Atomic diffusion on metal surfaces is a long term process that may induce undesirable topological modifications down to nanoscale, making these changes practically unnoticeable on large experimental surface areas which easily range from square micrometers to square centimetres. Understanding diffusion processes, including surface diffusion, becomes particularly important when dealing with applications that demand high technological precision (⩽ 1 μm), such as the components of accelerating structures of the future Compact Linear Collider (CLIC) [1]. In CLIC, the accelerating structures are designed to operate for extensive times under high gradient electromagnetic fields, which present additional challenges for keeping the metal surfaces unmodified. For instance, surface diffusion enhanced by an electric field is believed to induce nanoscale surface roughening on copper parts of the accelerating structures. The roughening leads to uncontrollable appearance of local vacuum discharges, damaging the surface and increasing the power consumption, thus decreasing the efficiency of the accelerator [2,3].

It is important to note that surface diffusion may play a crucial role also on a nanoscale, in the process of shaping of growing nanoparticles. For example, in [4], we showed that in a magnetron sputtering inert gas condensation chamber, iron nanoclusters grow cubic or spherical depending on sputtering intensity through the competition between surface diffusion and atom deposition.

The evolution of surfaces is even on a nanoscale a long-term process, not easily accessible by many existing simulation models. The kinetic Monte Carlo (KMC) method was specifically developed to simulate slow diffusional processes, which take place while the system evolves towards the potential energy minimum. Unlike other Monte Carlo methods, KMC is not only able to capture the ground state of thermodynamic equilibrium, but also able to estimate sufficiently well the kinetic path and the required time of a system to move towards the ground state [5]. The latter is enabled through the residence time algorithm [6], which estimates the time needed to complete a single transition.

The physics behind the KMC model is described by the probabilities of diffusion transitions. These probabilities can be estimated via transition energy barriers. Thus, a successful KMC model relies on appropriate estimation of the energy barriers of all possible transitions in the system. The most accurate methods, thus far, involve calculations of the barriers on the fly using the dimer method of finding potential transition paths on the potential energy surface [7] or applying self-learning procedures during the simulation [8–10]. Such methods usually operate off-lattice, allowing the inclusion of a large variety of possible transitions in the system, and require heavy computational resources. It is also common to use more simplified approaches for estimating the barriers, such as the approach of counting broken and newly forming bonds (the bonds before and after the transition) [11–13]. Such methods are less time consuming and easy to implement but they inevitably increase the uncertainty of the simulation results. Sophisticated mathematical techniques have been recently applied to calculate the energy barriers. Among them are cluster expansion [14,15], genetic programming [16], and artificial neural network [17–19] approaches. These methods are used to predict the energy barriers based on the local atomic environment. In [3], we presented the atomistic KMC (AKMC) model Kimocs for metal surfaces, in which we predefine the allowed transitions in the system and calculate the sets of energy barriers in advance.

---

\* Corresponding author.
*E-mail address:* ekaterina.baibuz@gmail.com (E. Baibuz).

Kimocs was designed to simulate evolution of nanofeatures on metal surfaces. It is clear that e.g. molecular dynamics (MD) methods are able to describe similar processes more accurately, since all the atomic configurations, which the system may have while evolving towards the energy minimum, occur naturally in MD. However, the limited time scale of MD methods does not allow to obtain any appreciable changes of a surface morphology with significant features and at temperatures well below the melting point. KMC, on the other hand, offers the possibility to reach rather long time scales with reasonable computational costs, provided that all the atomic jumps are described within the rigid lattice framework.

Since Kimocs was developed with the aim of simulating the temporal evolution of large nanofeatures, it was crucial to employ a parameterization scheme, which is both efficient and sufficiently accurate. We adopted the rigid lattice approximation, which describes an atomic system with all atoms occupying well defined positions in a given crystal structure. Within the rigid lattice approximation, a local atomic environment can be described by a finite number. Although the rigid lattice approximation has inevitable intrinsic limitations (for example, if surface reconstruction is expected to occur during the process simulated by KMC on a rigid lattice, such process will not be taken into account), it is an efficient approximation to develop fast algorithms that are minimizing computational costs.

In Kimocs, we constrain the transitions in a system to atomic jumps into vacant lattice sites, which we will henceforth call vacancies. The jumps may happen on the surface as well as in the bulk.

The use of a rigid lattice and the limitation on the variety of transitions make it possible to precalculate the sets of barriers for each material (Cu and Fe in this work). Precalculation of the barriers allows us to reach the desired efficiency of the simulation algorithm that only needs to assign tabulated barrier values to atomic jumps in this case. To assure the accuracy of calculated barriers we use the Nudged Elastic Band method. Although such a parameterization scheme seems to be straightforward and easy to implemented, we faced a number of challenges, which are difficult to circumvent.

In this article, we will focus on the calculations of energy barriers for AKMC models with a rigid lattice. We will discuss the challenges of the rigid lattice parameterization and how these challenges can be overcome in order to precalculate the migration energy barrier sets. We will present the tethering force approach, which allows to create heavy complete sets of barriers for all possible transitions on a rigid lattice. We are using this approach together with the parameterization scheme of the Kimocs model, but it is applicable for any other parameterization scheme in a rigid lattice, where possible transitions are restricted to a certain type, e.g. first nearest neighbour jumps in face-centred cubic (FCC) lattices (as is the case in Kimocs).

The structure of this paper is as follows. In Section 2, we provide some details of our KMC model Kimocs and parameterizations that have been used with it earlier. In Section 2.3 we describe the challenges that arise when migration barriers are calculated on a rigid lattice of FCC and BCC structures. In Section 2.4 we present the possible solutions to circumvent the problems described in Section 2.3 and introduce the tethering force approach (Section 2.4.2), which allows for calculations of the barriers on semi-rigid lattice, restricting the freedom of surface atoms to relax far away from they positions on a rigid lattice. In Section 3, we present different sets of migration barriers and discuss the limitations of each set along with the limitations of the Kimocs parameterization approach in general. In Section 3.2 we concentrate specifically on the sets where tethering is used and how this approach affects the KMC simulation results. Finally, we summarize our conclusions in Section 4.

## 2. Methodology

### 2.1. Atomistic Kinetic Monte Carlo on a rigid lattice

Before describing the challenges, which we encountered during the parameterization of our AKMC code Kimocs [3] for simulations of surface diffusion processes, we will briefly outline the basic principles of a rigid lattice AKMC model in general and describe in detail special features of our Kimocs code. In an AKMC algorithm within the rigid lattice approximation, a diffusion process proceeds via atomic jumps to a neighbouring vacancy. The event, which includes the choice of an atom to jump and the jump itself, is selected randomly, but with respect to the magnitude of the corresponding transition rates, which are compared for all events. This way, more probable events occur more frequently. The transition rates for all events in the system are calculated according to the Arrhenius formula for thermally activated processes:

$$\Gamma = \nu \exp\left(\frac{-E_m}{k_B T}\right), \tag{1}$$

where $\nu$ is the attempt frequency for the transition to occur, $k_B$ is the Boltzmann constant, $T$ is the temperature of the system, $E_m$ is the migration energy barrier, which the atom needs to overcome in order to move from one lattice site to another. For simplicity, $\nu$ is considered to be the same for all the transitions.

In Kimocs, the possible jumps in the system are restricted to primarily 1nn jumps in FCC and BCC materials, but 2nn jumps may also be allowed if necessary.

We precalculate the full set of the migration energy barriers, $E_m$, for all possible 1nn (and 2nn for BCC) jumps in the system to reduce the computation costs of simulations. The parameterization of the $E_m$ barriers is done within the 1nn and 2nn shell. Taking into account only the atoms in the nearest neighbourhood would result in insufficient accuracy, since the interaction with the atoms in the 2nn position is also quite strong in both FCC and BCC lattices.

Using both 1nn and 2nn shells in the parameterization scheme allows us to reach higher accuracy but leads to the full 26 (20 in BCC) neighbouring atoms description, which we will further refer to as the 26D parameterization scheme. In this scheme, if all barriers are to be calculated, then even in a mono-elemental metal $\sim 2^{26}$ barriers are needed.

The original more approximative parameterization scheme of Kimocs uses only four parameters to describe events and we will therefore refer to it as the 4D parameterization scheme. Within this scheme, each jump event is represented by four numbers, $(a, b, c, d)$, listed in the order of the number of 1nn ($a$) and 2nn ($b$) atoms of the initial configuration of the jumping atom and the corresponding numbers ($c$ and $d$) for the final vacant lattice site. The number of neighbours of the final vacancy is counted by assuming that the jump has not yet occurred, that is the jumping atom itself is included in $c$. It could also be illustrated as $a, b \to c, d$. A single permutation (a possible arrangement of neighbouring atoms and vacancies of the jumping atom, see example in Fig. 16) is chosen to represent all possible permutations for the given combination of the numbers $(a, b, c, d)$. Thus, the corresponding migration barrier of such an event, $E_m(a, b, c, d)$, is a function of four parameters. An example of an event description in the 4D approximation is shown in Fig. 1. Such an approach significantly reduces the set of necessary barriers down to ~5000 for FCC and ~2000 for BCC lattice structures.

If for some reason no barrier could be calculated for a particular $(a, b, c, d)$ combination, Kimocs assigns to this process a default near-infinite value (100 eV has been found to be enough), which will give a near-zero probability for the process to occur. This way we can forbid some improbable processes to occur.

The time evolution is an important aspect of a KMC model. The parameter that affects the time predicted by the calculated transition rates is the attempt frequency. In our parameterization scheme, we fit the attempt frequency to the MD simulations by comparing the flattening time of a surface nanotip as calculated by both KMC and MD (see details in [3]). In this manner, we obtain a sufficient accuracy for the time scale of our KMC simulations.

### 2.2. Nudged Elastic Band method

A key importance for any KMC model is the accuracy of migration energy barriers, which would allow to follow the correct kinetic path of the system towards the equilibrium. For systematic calculation of the actual values of the energy barriers for atomic jumps, we used the nudged elastic band (NEB) method [20–22] with semi-empirical potentials. The general algorithm we used for the calculation of a migration energy barrier with the NEB approach can be summarized as follows:

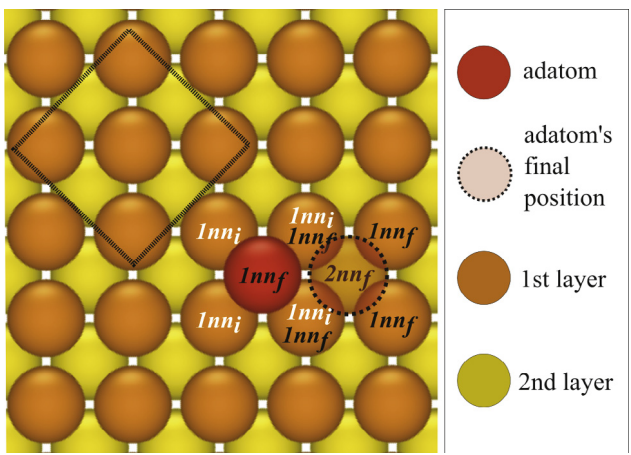

**Fig. 1.** Illustration of a (4,1,5,1) 1nn jump on a {100} FCC surface in Kimocs. Two surface layers are shown: top layer with orange circles and the layer below with yellow circles. The adatom (red circle) performs a jump from the site with four 1nn atoms and one 2nn atom (the atom right below the jumping atom) to a site (dashed semi-transparent circle) with five 1nn (including the jumping atom itself) and one 2nn atom below it (marked with $2nn_f$). The number of neighbours of the final vacancy is counted with the assumption that the jump has not yet occurred, that is, the jumping atom itself is included in $c$. To guide the eye, the FCC unit cell is shown with a square. (For interpretation of the references to colour in this figure legend, the reader is referred to the web version of this article.)

1. Relaxation stage: during this stage both the initial and final configurations of the event are relaxed towards the energy minimum.
2. Initial guess of the minimum energy path (MEP): usually, as an initial guess of MEP, the interpolation path from the relaxed initial position to the relaxed final one is chosen to be a straight line.
3. Relaxation of the interpolated path: initial energy path is relaxed towards MEP by using the NEB algorithm with MD and semi-empirical potentials.

For all the barrier sets for Cu surfaces mentioned in this article, we used the interatomic potential based on the Corrected Effective Medium Theory (CEM), developed by Stave et al. [23]. The potential describes well the properties of Cu surfaces [24]. The surface energies given by the CEM potential are in good agreement with both DFT calculations and experiments. The EAM potential developed by Mendelev et al. [25] was used for the barriers to describe the diffusion processes on Fe surfaces. Mendelev potential predicts well the general trend of surface energies of different surface orientations and vacancy formation energy [26], which was important for the studies in [4].

The energy barrier is found from the relaxed MEP as a difference in the potential energies of the initial configuration and the configuration at the saddle point. The different examples of MEP on potential energy surface profiles obtained with the NEB method are shown in Fig. 2. In the case of a symmetric jump (Fig. 2a), the potential energies of both the initial and final configurations are the same, thus the forward and reverse jumps have identical barriers and both events cause no change in the total potential energy. Such a case is rather specific and mostly, the barriers are asymmetric with respect to the potential energies in the initial and final configurations. If the final configuration has lower potential energy, the barrier towards it is also lower than the reverse one (see Fig. 2b). These barriers ensure that the more energetically favourable states are prioritised and the whole system evolves towards the potential energy minimum.

The barriers shown in Fig. 2a and b illustrate the processes that do not pose any problems in calculations with NEB. In this work we would like to focus on less clear situations, such as shown in Fig. 2c and d, which appear unavoidably during surface diffusion in a rigid lattice.

### 2.3. Challenges in barrier calculations on a rigid lattice

The situations in Fig. 2c and d are not trivial to interpret due to the absence of a clear saddle point on the MEP. Fig. 2c presents the case of spontaneous relaxation of an atom from initial ($l = 0.0$ in the figure) to final configurations ($l = ~2.5$ Å) in a rigid lattice during the NEB calculations. In a KMC algorithm, such a process should be assigned a small barrier to ensure that it happens with a high probability and within a short time. In [3], it was proposed to avoid the exactly zero barriers and instead use the following heuristic formula,

$$E_m(a, b, c, d) = \epsilon a + \delta b + \epsilon c^{-1} + \delta d^{-1} \qquad (2)$$

where $\epsilon = 10^{-3}$ eV and $\delta = 10^{-4}$ eV. This formula is designed to prioritise the jumps of atoms with the fewest neighbouring atoms. It also assumes that it is more favourable for an atom to jump into a position with a higher number of neighbours. $\epsilon$ and $\delta$ are chosen so that the number of 1nn atoms contributes more into the value of migration barriers than the number of 2nn atoms.

The MEP in Fig. 2d has an even more complicated shape, making it impossible to find a barrier from it. The minimum that occurs between the initial and final configurations along the MEP show that there are energetically favourable positions, which are off the lattice sites, but the rigid lattice approximation prevents atoms from occupying them. For this kind of jumps, it is not straightforward whether such processes have to be completely excluded from the simulations, or allowed to happen with some probability.

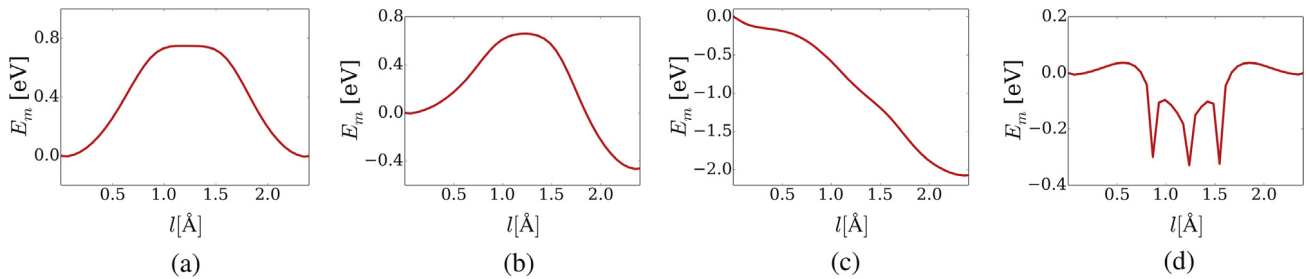

**Fig. 2.** Illustrative examples of minimum energy paths on potential energy surfaces for (a) a symmetric process; (b) an asymmetric process (c), a spontaneous process and (d) a process with several intermediate minima. $l$ is the distance from the initial position of the jumping atom; $E_m$ — potential energy (the energy reference is the potential energy at the initial position).

It is clear that such situations may happen due to a conflict of using the rigid lattice and an attempt to take into account strong relaxation effects driven by surface tension. For instance, when the evolution of a surface is considered in KMC algorithms via atomic jumps to 1nn vacancies in well-defined lattice sites in a rigid lattice (as in our Cu studies), many configurations, which are possible due to surface relaxation effects, are not accessible. In some cases, the atoms might be inevitably forced into configurations, which are energetically unfavourable and thus, unstable. These may not occur in a real diffusion process during dynamic simulations, however, in the rigid lattice, these unstable positions may be necessary, as they provide the only path for a system to reach the potential energy minimum state. In other words, if atoms are forbidden to occupy unstable positions in a rigid lattice, it may lead a system to the unphysical "frozen" state with no stable positions available for atoms to jump to. If so, the equilibrium may never be reached. To ensure that the system continues evolving, jumps involving atoms in unstable initial or final positions, or even both of them, should also be available in a rigid lattice, although these should be assigned the barriers that are estimated using some other technique.

More specifically, we noticed that atoms in the initial or final configurations of many jump processes on rigid lattice FCC surfaces may move and change lattice sites during the relaxation stage of the barriers calculations. In other words, the event $(a, b, c, d)$ in the 4D parameterization scheme becomes the event $(a', b', c', d')$ after the relaxation of the initial and final configurations and if the NEB calculations are then carried out, the barrier obtained in these calculations clearly cannot be used to describe the intended $(a, b, c, d)$ event. We call such configurations that change during the relaxation stage as *unstable*.

An example of an unstable configuration is shown in Fig. 3a. The figure shows the configuration, which is initial for the event $(2,1) \rightarrow (8,1)$ and final for the reverse event $(7,1) \rightarrow (3,1)$ (Note, that we count the number of neighbours of the initial atom and final vacancy, i.e. a jumping atom is included in the number of 1nn of a final configurations, $c$, in both cases). In Fig. 3b, we show the configuration which is final for the event $(2,1) \rightarrow (8,1)$ and initial for the $(7,1) \rightarrow (3,1)$ event. Already during the initial relaxation stage, we noticed that the configuration in Fig. 3a turned into the configuration in Fig. 3b (the jumping atom shown with blue relaxed one monolayer down towards the higher number of neighbours). Thus, we recognize the event $(2,1) \rightarrow (8,1)$ as *spontaneous* and assign it a small near-zero barrier to ensure that this process happens at once when the initial configuration $(2,1)$ of the process $(2,1) \rightarrow (8,1)$ occurs during KMC simulations. A spontaneous process can also be detected directly from the results of the NEB calculations: if the initial position has higher potential energy than the final position and the saddle point does not appear during the NEB relaxation towards the MEP (Fig. 2c).

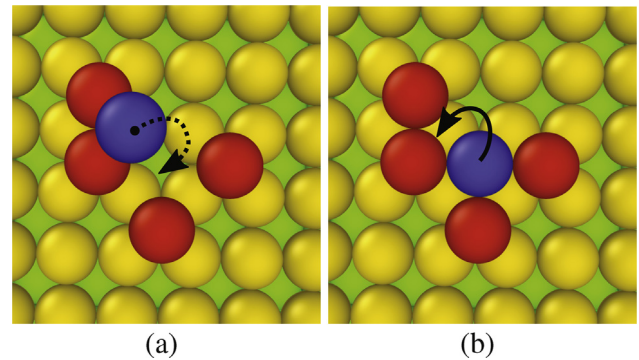

**Fig. 3.** Top view of (a) the unstable initial configuration of the process $(2,1) \rightarrow (8,1)$ (final configuration of $(7,1) \rightarrow (3,1)$), (b) the initial configuration of the jump $(7,1) \rightarrow (3,1)$, which is also final of $(2,1) \rightarrow (8,1)$; yellow colour corresponds to surface atoms, red - adatoms, the jumping atom is shown with blue; in (a), the jumping atom is one monolayer above the adatoms, in (b) - on the same level; a full arrow shows a jump to one monolayer up; a dashed arrow - to one monolayer down. (For interpretation of the references to colour in this figure legend, the reader is referred to the web version of this article.)

On the other hand, the reverse event $(7,1) \rightarrow (3,1)$ in Fig. 3b should be assigned a high barrier so that this process will seldom take place in the simulations.

Such an assumption would allow to include (at least indirectly) the transitions to further than 1nn vacancy as a part of a multi-step transition. Thus the atoms, which were forced to the physically unstable positions during the first step, will relax during the next KMC step either back into the initial position or to another position, which is more stable. This issue will be discussed later with respect to the diffusion in BCC structures.

Moreover, the $(a, b, c, d)$ event can also be modified during the NEB relaxation steps, although both initial and final configurations of the process are stable. Since the NEB algorithm forces the system towards the energy minimum at every step (see Section 2.2) — in some cases with many vacant lattice sites around the atoms (open surface, large vacancy clusters, etc.) — it may also force either a jumping atom away from its initial path or the surrounding atoms away from their initial positions, confusing the calculation of the barrier for the initially determined configuration. Fig. 4 shows an example of a transition onto a void {1 1 1} surface, where the jumping atom (A) was dragged by the NEB relaxation procedure into the different final position (C) instead of the intended one (B), although both initial (A) and final (B) configurations of the intended process are stable.

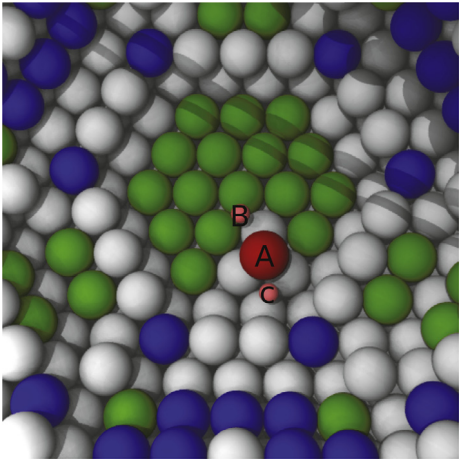

**Fig. 4.** Unstable process on the void {111} surface in FCC structure. Green atoms are part of the {111} surface, blue atoms are in the {100} surface, grey atoms are other surfaces. The jumping atom is red. It is set up to jump from the position A to position B, but instead it is dragged into C by NEB. (For interpretation of the references to colour in this figure legend, the reader is referred to the web version of this article.)

## 2.4. Possible solutions to circumvent uncertainties in barrier calculations on a rigid lattice

### 2.4.1. Simple model to avoid unstable configurations

In [3], we suggested a way to estimate the barriers of many unstable configurations. We assumed that all events with $a \leqslant 3$ (too few 1nn atoms) were spontaneous, since they are most likely to involve atoms in unstable initial configurations. These events were assigned barriers given by Eq. (2), regardless of the stability of the final positions. The maximum migration energy of processes with $a \leqslant 3$, which can be obtained with Eq. (2), is $E_m(3, 6, 1, 1) = 0.0047$ eV, which is insignificant compared to even thermal energies (>0.025 eV). Most of the jumps from stable configurations to unstable ones, as $(7,1) \rightarrow (3,1)$ in Fig. 3 along with more exotic configurations as in Fig. 2d, are forbidden in this approach. However, some of this kind of processes were manually identified to cause a "frozen" state during the nanotip simulations. If that was the case, we used simplified approaches to calculate barriers of these processes: in the case if the jumps were from stable configurations to unstable ones, we performed the NEB calculations with partial relaxation of the final positions (the surrounding atoms were allowed to relax, while a jumping atom was fixed); in the case of more exotic configurations as in Fig. 2d, a barrier was either chosen as a difference between a maximum on the MEP and initial state energy or marked as spontaneous if initial state energy was the highest point on the MEP.

The use of Eq. (2) for all the atoms with less than three neighbours is not fully justified, although it was motivated by the possibility to avoid artificial scenarios. In AKMC simulations, it becomes problematic, when a small cluster of atoms with many possible processes with essentially zero barriers appears on a surface. Atoms might very well jump away in such an order that one or more atoms are left behind without any neighbours. An isolated atom, which has a zero probability to make a jump, has to be either considered removed from the system (which easily leads to the destruction of the surface) or left isolated. Either of these options is artificial as the current parameterization does not include evaporation events. Eq. (2) allows to circumvent evaporation of clusters of atoms. Nevertheless, there is still a chance that many events with ($a \leqslant 3$) might have higher barriers than given by Eq. (2). This may lead to over-relaxation of the surface: some events treated as spontaneous, need a longer time in reality before they can take place.

Although, a sufficiently large amount of barriers was eventually calculated in [3] to overcome the problem of a "frozen" state, the method of calculations was not efficient and time consuming since many of the processes were treated manually. Also the number of remaining forbidden events was still not negligible. In the next section, we present a recently developed automatized approach to calculate the migration energy barriers within the rigid lattice approximation that allows to tabulate the whole set of possible $E_m(a, b, c, d)$ barriers and minimise the use of Eq. (2).

### 2.4.2. Tethering force approach

Since surface relaxation effects in some cases may cause modifications of initial or final configurations during the NEB calculations of energy barriers, we deployed tethering forces on atoms to attract them towards the initial lattice positions. Thus, atoms are positioned on a semi-rigid lattice, where they are allowed to relax but only within the proximity of their initial lattice positions. This approach was specifically designed to calculate the barriers for unstable configurations and appears to be a plausible and efficient solution to the problem at hand.

The automatized scheme for barrier calculations utilizing the tethering force approach can be described as follows. To calculate the barrier, the method selects a configuration out of all permutations corresponding to the same numbers $(a, b, c, d)$, which results in the lowest sum of the energies of the initial and final states. Since the choice of permutations is consistent, it is possible to obtain barriers for both $(a, b) \rightarrow (c, d)$ and the reverse $(c, d) \rightarrow (a, b)$ events from the same minimum energy path calculated by NEB. The energy of the system depends on the used interatomic potential, so the permutation selection process needs to be done for each potential separately.

After all configurations representing the families of different permutations are defined, the energy minimisation algorithm for initial and final positions is launched. As discussed above, some configurations, e.g. with too few neighbours, may be unstable. In our approach, we distinguish between the following different situations:

1. Both the initial and final positions minimise correctly. In this case, the NEB calculation proceeds.
2. The process is indicated as spontaneous either during minimisation or the NEB relaxation and assigned the migration energy according to Eq. (2). The reverse process is calculated as the energy difference between the initial and final states.
3. The jumping atom relaxes to some unrelated position from either the initial configuration or the final one. Alternatively, the neighbouring atoms move during the MEP calculations, resulting in a modified $(a', b', c', d')$ event compared to the intended $(a, b, c, d)$. In this case, we do not obtain the barrier for the $(a, b, c, d)$ event. Moreover, the new event $(a', b', c', d')$ cannot be identified either.

The processes described by the situation 3 are rather unlikely to happen on real surfaces, but inevitable in the rigid lattice approximation. These also present the biggest challenge in the calculation of the barriers. To address it, we use the tethering force approach. During minimisation, an additional spring tether is applied to all atoms in a simulation box. This spring attracts atoms to the lattice sites, where they belong. Depending on the strength of the spring tether, the atoms can deflect from their initial sites more loosely or more rigidly. In this manner, all the barriers, including the barriers for unstable configurations, may, in principle, be calculated.

Furthermore, we noticed that during the NEB relaxation, as the path of the jumping atom converges towards the MEP, the surrounding atoms sometimes also change their positions. These situations are particularly difficult to detect. Even though the initial and final states are relaxed (or tethered and relaxed), the potential energy landscape on the surface can favour different intermediate positions for the atoms. As a result, the energy values fluctuate along the path and extra energy minima occur (see Fig. 2d). It becomes unclear how to define a saddle point for such MEP. To reduce this effect, we apply tethering also during the NEB calculations. The spring in this case tethers the neighbouring atoms in each NEB image to their initial and final positions.

The tethering spring force complements the NEB spring force, which binds the atoms between images. The energy contribution from the tethering force is not included in the energy landscape and migration barrier calculation. However, the energy landscape is affected indirectly because tethering affects the positions of the surrounding atoms. In this way, the effect of the tethering force on migration barriers is minimized (see Section 3.2). Tethering is not used on a jumping atom, thus allowing it maximum freedom of movement to find the MEP. However, even with this approach there are processes with minimum energy paths similar to Fig. 2d. This issue will be addressed in Section 3.2.

Tethering force approach also allows to overcome the problem shown in Fig. 4, when a jumping atom is dragged by NEB into a position, which is different from the intended one.

We chose a constant value of the tethering force in such a way that it allows to increase significantly the number of barriers (including unstable configuration events) that can be calculated in a consistent manner. On the other hand, as the tethering force implicitly affects the migration barriers (see Fig. 11), it is advisable to use the lowest working force constant.

A number of stable and unstable configurations were chosen to test different tethering constants for copper with the CEM potential. Initial/final state minimisations and NEB calculations were performed with the tethering parameter values 0.0, 0.35, 0.5, 1.0 and 2.0 eV/Å². We found that all the values up to 1.0 eV/Å² are not sufficient to hold unstable atoms from relaxing to unrelated positions. With tethering parameter 2.0 eV/Å², all barriers were calculated successfully and the barriers for the event involving stable positions were also still reasonably close to the barriers calculated without tethering force (see Section 3.2 for details). Thus, the tethering parameter for Cu was chosen to be 2.0 eV/Å².

The tethering force approach helps to overcome the challenges introduced by a rigid lattice and calculate the barriers for many unstable configurations, for example in Fig. 5. The consistency of the tethering force approach allowed us to develop a framework for efficient barrier calculations in a rigid lattice. To produce a set of barriers for Kimocs simulations within the 4D parameterization scheme, the framework requires only an interatomic potential and the restricting parameter of how far away atoms can move from their initial positions (a 1nn distance for FCC materials). The method is not restricted to the use with Kimocs only, but can be generalized to any parameterization scheme in a rigid lattice.

In the next section, we will turn our attention to the different types of challenges that are especially important for the simulations of surface diffusion in BCC metals.

## 2.5. Specific challenges in parameterization of the {100} BCC surface and possible solutions

Parameterization of a KMC model for surface diffusion in BCC metals becomes challenging when a {100} surface is considered. The assumption that adatoms diffuse on the surface mostly via 1nn jumps is no longer valid. For instance, a jump of an adatom from a hollow site (on top of four surface atoms) to a neighbouring hollow site will be a 2nn jump (Fig. 6a). It has been previously seen that a diagonal exchange is one of the most probable diffusion processes on BCC {100} surfaces [27]. A diagonal exchange is a multi-step diffusion jump. In this process, an adatom dislodges a surface atom in a 1nn site and takes its place. After this, a surface atom is forced up to the surface (another 1nn jump), occupying a 3nn position with respect to the starting point of the first adatom, see Fig. 6e. Another possibility for the dislodged surface atom is to occupy a 2nn position with respect to the starting point of the initial adatom (Fig. 6b). However, this non-diagonal exchange process has much lower, but not completely negligible, probability than both a diagonal exchange and a 2nn jump [28]. In order to fully describe the diffusion on {100} BCC surfaces, all three types of processes should be included in the KMC algorithm: 2nn jumps to hollow sites, diagonal and non-diagonal exchange processes for all possible local atomic environments of the jumping atom.

In the rigid lattice description, atomic jump processes are characterized only by initial and final states of the system, the specific transition path from the initial to the final state does not play any role. Thus, we do not distinguish between the different transition paths resulting in the same initial and final states. In other words, the end result of single-atom jumps and concerted (exchange) jumps looks the same in Kimocs: like only one atom changed place. Diagonal and non-diagonal exchange processes will thus appear as effective 3nn and 2nn jumps, respectively, of the initial adatom in a rigid lattice (Fig. 6d and a, respectively). Thus, the rigid lattice parameterization scheme of BCC {100} surfaces should be extended to the 3nn coordination shell and include the distinction between the direct and the exchange jumps. This task turns out to be exceptionally demanding as the computational costs of a KMC algorithm increase with the increasing number of possible jumps in the system and as barriers must be calculated for all three types of jumps described above for various configurations.

Kimocs does not explicitly include the description of exchange processes, but an exchange process may effectively happen via two consecutive 1nn jumps: first, a surface atom below the jumping adatom breaks out of the surface layer and becomes an adatom itself. Then, the two adatoms with equal probability can fill in the formed vacancy in a spontaneous jump (see Figs. 6c and f). For this to occur, the surface atom must first overcome a barrier to occupy the 2nn or 3nn position with respect to the "jumping adatom". If these barriers are calculated directly by NEB for either process (to 2nn or 3nn positions), they are much higher than the barriers calculated for the processes of the dislodging of the surface atom by the jumping adatom. One of the solutions, which can be suggested to address this problem, is to lower the barrier artificially for those surface atoms which are close to the adatoms, but in that case artefacts may appear for some other permutations corresponding to the same numbers describing the event. At the same time, the direct 2nn jump can easily fit within the 4D parameterization scheme adopted in Kimocs. This only requires an additional step to distinguish between 1nn and 2nn jumps, while the event description numbers $(a, b, c, d)$ can be used for both.

In the diagram shown in Fig. 7 we summarise the challenges, which arise from the use of a rigid lattice approximation to simulate surface diffusion processes and the solutions which can be used to circumvent the problems related to strong relaxation processes on the surface.

## 2.6. Sets of migration energy barriers

We used the approaches to circumvent the problems related to surface migration barrier calculations (see Section 2.4) in different Kimocs parameterization tables, referred hereafter as Sets. Here we present and analyse four sets of barriers: *Cu Set 1*, *Fe Set 1*, *Cu Set 2* and *Cu Set 3*. The full sets of copper and iron barriers are submitted along with this article as Data in Brief entries and can be found in [29,30], respectively. The sets can be used for simulations with Kimocs or other similar codes, which parameterize atomic jumps in the rigid lattice with the number of neighbours of a jumping atom before and after the transition (e.g. [31,32]). For the reader's convenience, we summarize below the ways of how the different sets were calculated and give more technical details.

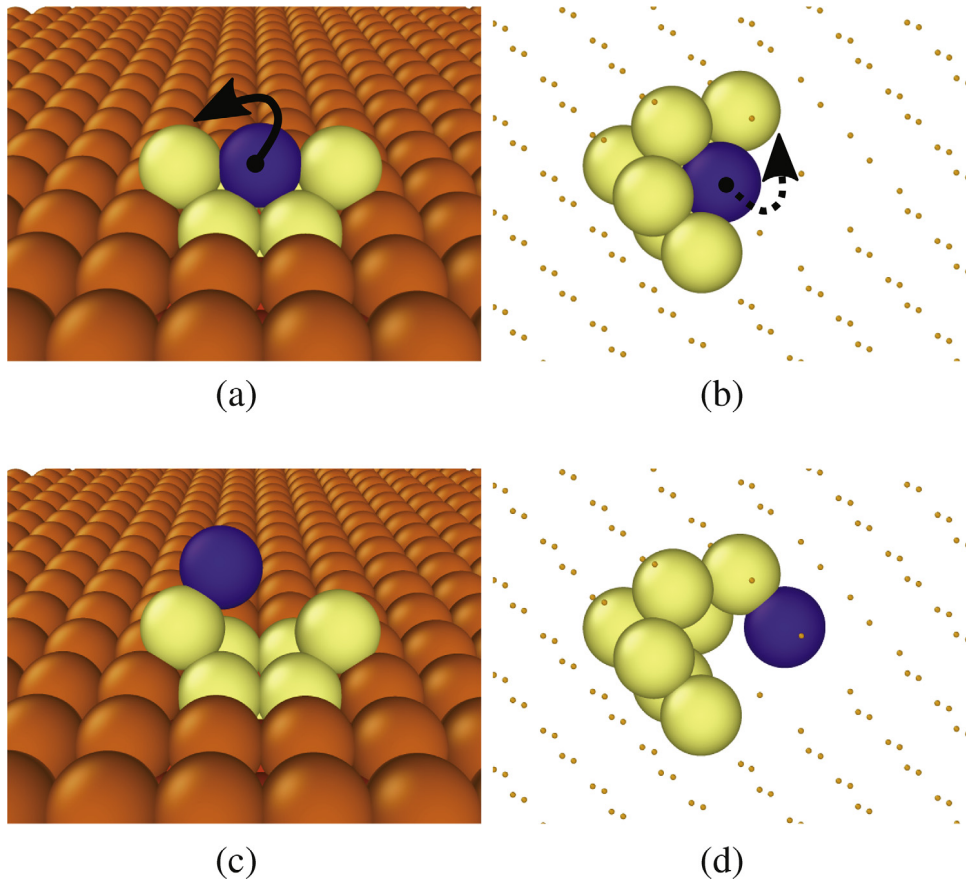

(a)

(b)

(c)

(d)

**Fig. 5.** Example of a configuration on the surface that required the tethering force to enable the calculation of the barrier of an atom shown with blue colour to jump from the position (a) to the position (c). (b) and (d) show the arrangement of atoms corresponding to the same process but inside the bulk. 1nn and 2nn atoms of the jumping atom before and after the jump are shown with yellow colour. Orange atoms correspond to those outside the 2nn shell of the jumping atom (shown with small dots in the bulk pictures). Arrows emphasize that an atom jumps to the 1nn positions one monolayer up (full arrow) or down (dashed). (For interpretation of the references to colour in this figure legend, the reader is referred to the web version of this article.)

### 2.6.1. Cu Set 1

The first set, *Cu Set 1* [29], was calculated as described in Section 2.4.1. More specifically, we did not apply a tethering force or other restrictions during the NEB calculations and some barriers were identified as spontaneous, while some were forbidden. *Cu Set 1* was successfully used to simulate the Cu surface self-diffusion in [3]. In this table, permutations were defined at random on {100}, {110}, {111} surfaces or a bulk system with a random distribution of vacancies. This set includes the barriers for 4289 $(a, b, c, d)$ events, most of which were calculated in the bulk. Among these, 2486 events with $a \leqslant 3$ and only 190 events with $a > 3$ were classified as spontaneous and assigned the barrier values according to Eq. (2). The NEB calculations were done using the MD code PAR-CAS [33–35]. We used the approach described in [36] for the calculation of the additional NEB spring force between the images. A sequence of 40 images was used for every jump. The initial and final images were relaxed with the conjugate gradient method and then fixed during the NEB calculations. The attempt frequency was fitted to MD simulations of the flattening time obtained for Cu surface nanotips (see [3] for further details). The obtained value of the attempt frequency, $v = 7 \cdot 10^{13}$ s$^{-1}$, compares well with the Debye frequency for Cu, $v_D = 4.5 \cdot 10^{13}$ s$^{-1}$ [37–40].

### 2.6.2. Fe Set 1

The second set, *Fe Set 1* [30], includes 1760 barriers, most of which were calculated in the bulk, for 1nn jumps with 214 barriers assessed by using Eq. (2). Barriers for jumps on the {100} and {110} surfaces were calculated separately and are prioritised in the set. The same scheme that was used for *Cu Set 1* was utilised for *Fe Set 1* as well, although here we did not assume the events with too few neighbours in the initial configuration to be spontaneous. Instead, all the barriers for the range of events with $a \leqslant 8$ were calculated directly. The table of 1nn jumps was complemented with the 16 $(a, b, c, d)$ barriers for the direct jumps of adatoms to a vacant 2nn site on the {100} surface (these barriers for 2nn jumps are combined to *Fe Set 2NN* [30]). *Fe Set 1* and *Fe Set 2NN* have been used in [4], where the attempt frequency for all events was assumed to be $v_D = 6 \cdot 10^{12}$ s$^{-1}$ [41]. In [30], we also provide the set of exchange processes, *Fe Set Exchange*, on the Fe {100} surface.

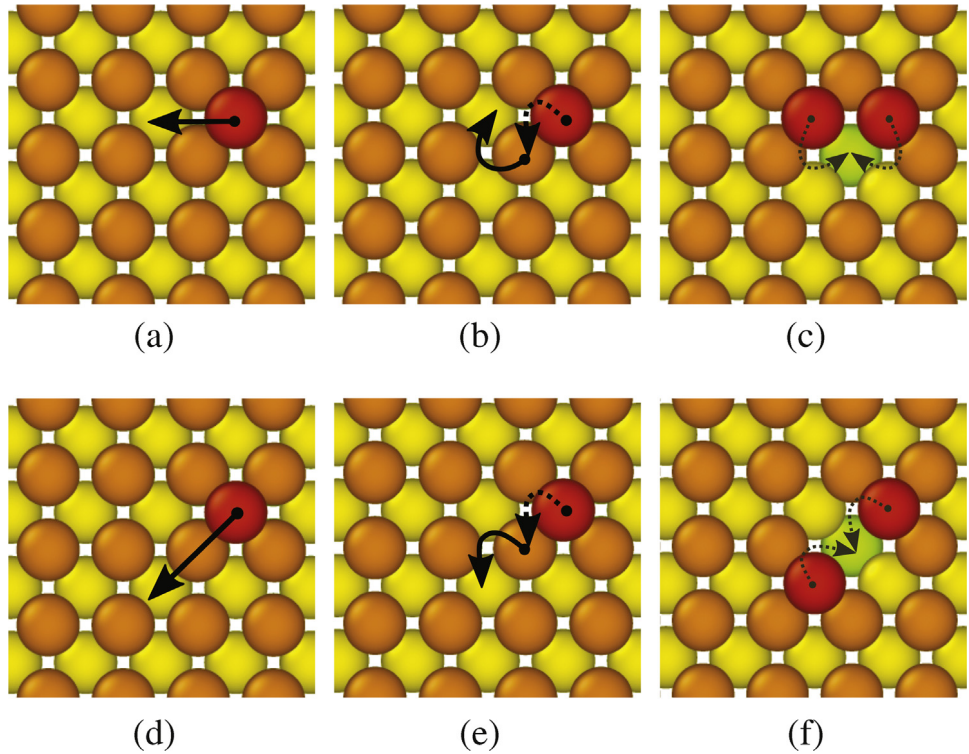

**Fig. 6.** Illustration of the 2nn (non-diagonal, (a)) and 3nn (diagonal, (d)) jump on the BCC {100} surface, which may take place effectively via an exchange process on a rigid lattice. (b) Illustrates the completion of the non-diagonal jump and (e) the completion of the diagonal jump; (c) and (f) show two possible 1nn jumps that might follow the first 1nn step of the non-diagonal and the diagonal exchange process in Kimocs (see the main text for details). Adatoms are shown with a red colour; straight arrows show jumps within the same monolayer; full curved arrows show jumps to one monolayer up; dashed curved arrows show jumps to one monolayer down. (For interpretation of the references to colour in this figure legend, the reader is referred to the web version of this article.)

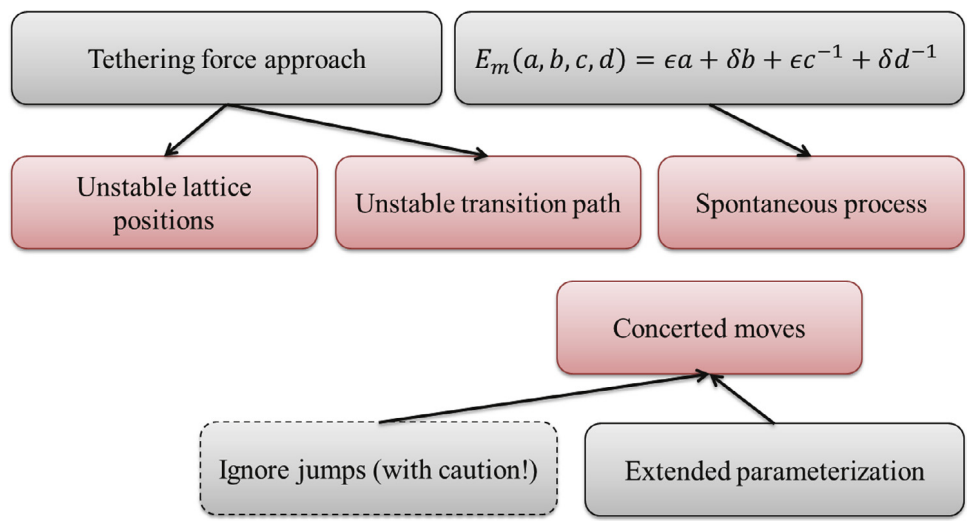

**Fig. 7.** Diagram of challenges (in pink) and solutions (in grey) discussed in this work. The tethering force approach helps to overcome problems of stabilization both in initial and final positions and in transition paths of the jumping atom. For spontaneous processes, Eq. (2) is used. To take account of concerted moves, a different parametrization scheme would have to be developed. (For interpretation of the references to colour in this figure legend, the reader is referred to the web version of this article.)

### 2.6.3. Cu Set 2 calculated with the tethering force approach

*Cu Set 2* [29] was generated utilizing the tethering force approach, as was discussed it Section 2.4.2, and was thus generated in a more rigorous and systematic way than *Cu Set 1*.

Here, the $E_m(a, b, c, d)$ barriers were first calculated for the hop-on jumps of an adatom on {100}, {110} and {111} Cu surfaces as we were interested in these surface orientations for simulations of nanotips. By hop-on jumps of an adatom we henceforth mean the events where the adatom jumps within the first layer above the surface having its 1nn and 2nn atoms constrained to the same layer of adatoms and below. For configurations $(a, b, c, d)$ that were available on more than one surface, the barrier for the {111} surface was prioritised, followed by the {100} surface, with the {110} surface as the lowest priority. This ordering is based on the surface energies for Cu [42], with the lowest surface energy taken as the most important. Barriers that were not available on any surface, i.e jumps with a large number of neighbouring atoms in the initial and final configurations that included many atoms in unstable positions on the surface (see an example in Fig. 5c and Fig. 3) were calculated in the bulk. Clusters of vacancies were built around a jumping atom in the bulk (see an example in Fig. 5), which was periodic in the x and y directions, but not in the z direction. The vacancies were created so that the number of neighbours of the jumping atom corresponded to the given $(a, b, c, d)$ combination in order to mimic the surface $(a, b, c, d)$ event, since the permutations of vacancy positions are not taken into account in the 4D parameterization scheme approach. This set of barriers, calculated using the tethering force approach, we will henceforth refer to as *Cu Set 2*.

In *Cu Set 2*, the barriers were calculated both on a surface and in a bulk. The Lammps MD package [43] was used for the NEB calculations with a climbing image [22] and an additional tethering force, with the tethering force constant set to 2.0 eV/Å$^2$, was applied (see Section 2.4.2) and a total of 24 images were used. The energy barriers for 5103 $(a, b, c, d)$ events were calculated in *Cu Set 2*. Only 211 barriers of jump event with $a \leqslant 3$ were assessed with the use of Eq. (2); the rest of the barriers were calculated with the tethering force approach. An additional 72 processes were marked as spontaneous for processes with $a > 3$. None of the jumps are forbidden in the tethering force approach, thus *Cu Set 2* is complete. The attempt frequency value $v = 3.1 \cdot 10^{14}$ s$^{-1}$ was again found by fitting of tip flattening time to MD results as described in [3]. *Cu Set 2* was used to reproduce the results of [3], which will be discussed in Section 3.2.

### 2.6.4. Cu Set 3

*Cu Set 3* [29] was calculated within the 26D parameterization scheme, which takes into account permutations of the neighbours of the jumping atom.

For *Cu Set 3*, we first calculated the barriers with many neighbours in the bulk for 334,725 random permutations of random $(a, b, c, d)$ configurations. We noticed that some of the bulk events resulted in unphysically high barriers. Inspection of such situations showed that these barriers resulted from the negative surface stress on the void interface contracting the entire simulation box during the NEB relaxation. This problem motivated the return to the original method to calculate the barriers with too high values (>1.5 eV) on the surface. The same 1nn and 2nn local atomic environment of the jumping atom was constructed close to the surface as it was in the bulk (see an example in Fig. 5). A tethering force constant of 2.0 eV/Å$^2$ was sufficient to hold the atoms in their positions and only one barrier calculation failed out of more than 330,000.

*Cu Set 3* is incomplete, thus we do not recommend this set for any physical simulations. For a complete 26D set, 8 million barriers would need to be calculated, which is infeasible for storage and handling during the present AKMC simulations by Kimocs. A special way to link such a large 26D set to the AKMC simulations has to be designed. Nevertheless, *Cu Set 3* is very useful for the analysis advantages and disadvantages of the 4D versus the full 26D parameterization schemes, used in other above-described sets.

## 3. Results and discussions

In this section we will focus on the discussion of reliability of different approaches chosen to calculate the surface migration energy barriers described above. For illustrative purpose, we plot the migration barriers for each set as a function of the difference of the number of 1nn atoms in initial and final positions, $a - c$, in Fig. 8. It is clear that many energy barriers may correspond to the same $a - c$ value. However, we can follow the variation of the barriers as a function of a gradual change of the situations: the negative values, $a - c < 0$, show barriers for the jumps from less to more stable positions (more neighbours in the final state) and the positive ones, $a - c > 0$, show barriers for the jumps from more to less stable positions (less neighbours in the final state). The colour scale corresponds to the amount of events which can be described by the same $a - c$ and $E_m$ values.

### 3.1. Kinetics described by the non-tethered sets

At first, we show that even the barriers of *Cu Set 1* and *Fe Set 1*, which were calculated with the initial simplified schemes, where surface relaxation effects were approximated via events with either near-zero barriers ("spontaneous" events) or very high barriers ("forbidden" events), are able to capture the physics of surface evolution via diffusional processes on copper and iron surfaces, respectively.

We have previously shown that the results of the KMC simulations of the stability of Cu surface nanotips [3] and formation of Fe nanocubes [4], where the sets were used, agreed very well both with MD simulations and experiments. However, in the current paper we still take a closer look into the kinetics of surface events, which may not be crucial for the previous results, but still can be overlooked while using *Cu Set 1* and *Fe Set 1*.

#### 3.1.1. Analysis of the Fe Set 1 barriers

The abundance of the calculated barriers for *Fe Set 1* can be seen in Fig. 8a. Some of the barriers (for no more than 300 events) in this set were calculated on the surface of a nanowire in order to emulate the conditions (a combination of a certain deposition rate and a temperature), where the formation of nanocubes was observed. It was found that the growth of Fe nanoparticles into cubic shapes is due to the difference in the rates of atomic jumps on {100} and {110} surfaces, which becomes more significant at low temperatures.

As the analysis of the processes happened during the nanocubes formations showed, most of the $(a, b, c, d)$ jumps were hop-on processes on {110} and {100} surfaces. Other processes correspond to the deposition events. Thus, neither barriers calculated in the bulk nor spontaneous processes calculated with Eq. (2) played a crucial role — in contrast to the barriers calculated on {110} and {100} surfaces for stable hop-on events.

The hop-on jumps of an adatom surrounded by various numbers of 1nn and 2nn, located only in the layer of adatoms, on the {110} surface have a wide range of barriers from 0.1 eV to 1.1 eV (Fig. 9). The low barriers correspond to the simple hop-on jumps with few neighbours and the jumps towards the adatom islands. High barriers correspond to the detachment of adatoms from adatom islands. On the {100} surface, the diffusion is driven by the 2nn jumps and exchange events, as was discussed in the Section 2.5. These processes have barriers higher than 0.6 eV (for simplicity's sake, non-diagonal exchange processes on {100} surface are not included in Fig. 9, because of much higher barriers than 2nn and diagonal exchange jumps on Fe BCC {100} surface [28]). Thus, at low temperature conditions (below 1000 K), where nanocube formation was observed in experiments, diffusion of single adatoms and small nano-islands on the {110} surface is much faster than on the {100} surface. On the other hand, at high temperatures, events on the {100} surface become as probable as jumps on the {110} surface. Nanoparticles may thus grow into a close-to-spherical shape driven by the surface minimisation at high temperatures.

As it was pointed out in Section 2.5, the exchange processes were not explicitly added to the KMC algorithm. Instead, the exchange events were treated in two steps as shown in Fig. 6. The first step of the 3nn effective jump corresponds to the 1nn $(5, 5, 4, 1)$ jump event, which is the same as for a vacancy jump inside the bulk close to the {110} surface. On the {100} surface, the $(5, 5, 4, 1)$ event has a barrier 0.72 eV, whereas a vacancy jump close to the {110} surface has a barrier of 1.65 eV. In order to avoid artificial void formation inside the bulk, we had to use a higher barrier for the $(5, 5, 4, 1)$ event in *Fe Set 1*. If so, the 3nn effective jump, which takes place most frequently during adatom diffusion on the {100} surface, was almost inaccessible in our simulations. If such a high barrier is used, the difference between the KMC and MD simulations will still appear even at high temperatures, at which the rates of diffusion on {110} and {100} surfaces become comparable. On the other hand, the rate of the 3nn effective jump at low temperatures (<1000 K) is still much lower than that of hopping events on the {110} surface (the corresponding barriers are 0.72 eV and ~0.27 eV). Within this range of temperatures, our KMC simulations agree well with MD simulations.

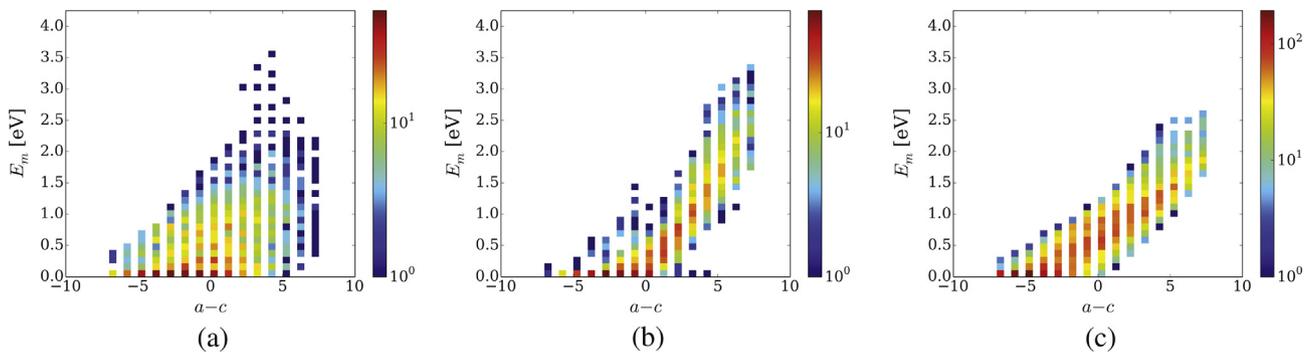

**Fig. 8.** Distribution of the migration energies vs. the change in the number of 1nn of the initial and final positions of the jumping atom for $(a, b, c, d)$ jumps in (a) *Fe Set 1*; (b) *Cu Set 1* ($a \leqslant 3$ processes are not included); and (c) *Cu Set 2*. colour corresponds to the occurrence of the migration energies for certain values $(a - c)$. For simplicity, $c$ does not include a jumping atom. (For interpretation of the references to colour in this figure legend, the reader is referred to the web version of this article.)

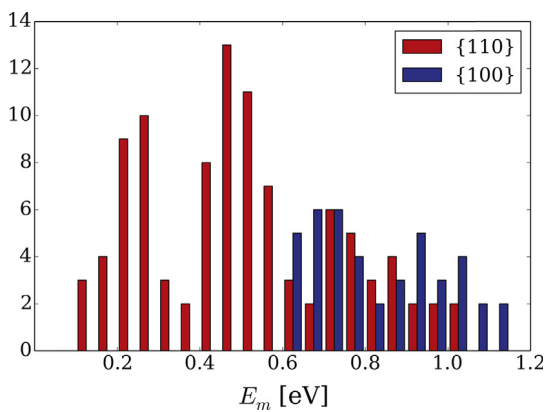

**Fig. 9.** Comparison between the diffusion events on {100} and {110} iron surfaces. A histogram for {110} surface includes only hop-on jumps, {100} includes the second nearest neighbour hop-on jumps and diagonal exchange processes of *Fe Set 2NN* and *Fe Set Exchange* in [30].

To verify that diagonal exchange jumps are not accessible at low temperatures also in MD, we calculated the barrier for this multi-step jump with different neighbourhood arrangements of the adatom's initial and final positions. The lowest barrier, which we found with the Mendelev potential, was 0.63 eV (see *Fe Set Exchange* in [30]), which is still much higher than 0.27 eV barrier of hop-on process on {110}.

Based on this analysis, we conclude that the overall difference in the diffusion rates on {100} and {110} surfaces can be well predicted already within the 4D parameterization scheme of 1nn and

2nn jumps at low temperatures. However, the exchange events must be included into the simulations more accurately to predict reliably the kinetics of events also at high temperatures.

### 3.1.2. Analysis of the Cu Set 1 barriers

Fig. 8b illustrates the abundance of the barriers in *Cu Set 1*. In this figure, neither forbidden nor spontaneous events assigned close to zero barriers are included. The trend of the plot indicates a clear monotonic growth from smaller barriers for $a - c < 0$ towards the higher barriers for $a - c > 0$. The barriers on the left side are mostly below 0.5 eV. The individual cases of relatively high barriers (blue dots on the top of each column) correspond to the unstable final configurations that were handled using the approach discussed in Section 2.3. The right part of the graph has most of the barriers greater than 1 eV and no barriers less than 0.5 eV, meaning that jumps from stronger bonded states are much less likely to happen. Such a trend is not true for iron barrier distribution (see Fig. 8a), which can be explained by the FCC materials being closer packed compared to BCC: it is harder for atoms to break the bonds in the stronger bonded position in the FCC structure in order to jump to a lesser bonded state ($a - c > 0$).

We also see in Fig. 8b that the red colour is more pronounced for the events with $a - c < 0$ where it is less probable that the final configuration of the atom is unstable. Thus, a smaller number of barriers in the right part could be explained by there being less combinations of $(a, b, c, d)$ processes, for which we were able to obtain the barriers with the approach described in Section 2.4.1. It is also interesting to note that there is a whole range of the barriers $(0, 1.5)$ for the $a - c = 0$ situations, which indicates the effect of the 2nn atoms on the value of the barrier. Those were included in the calculations, but are not shown explicitly in the plot.

Furthermore, we analysed the $(a, b, c, d)$ events that happened during the flattening of a cuboid nanotip of 13 nm height and 2 nm width constructed on a {110} surface, which was simulated using *Cu Set 1* (for details of the KMC simulations, see [3]). As we noticed during the flattening process, {111} facets were quickly built around the foot of the tip, allowing the adatoms to slide down from the top of the tip towards the surface. We analysed all the events, which took place during the flattening of the tip at 1000 K. These corresponded to a wide range of the energy barriers from near-zero (spontaneous) up to 2.0 eV (Fig. 10). Most of the events were below 0.2 eV, which are the transitions from lower to higher number of nearest neighbours. Spontaneous events happened only in 0.1% of the jumps. 1% of all the jumps corresponded to the events calculated in the bulk.

The analysis of individual events revealed the existence of a key event involved in flattening: adatoms on a {100} facet of the tip slide down to the {111} facet near the foot of the tip via the

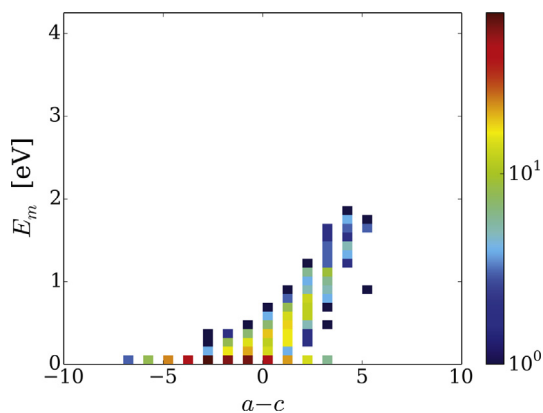

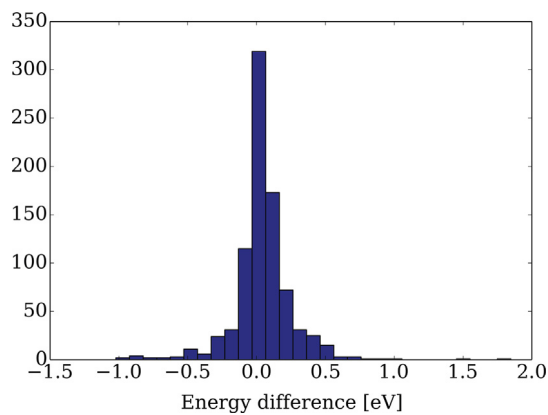

**Fig. 10.** Distribution of the migration energies vs. the change in the number of 1nn of the initial and final positions of the jumping atom for $(a, b, c, d)$ jumps during the flattening of a Cu cuboid nanotip of 13 nm height and 2 nm width, constructed on a {110} surface at 1000 K. Colour corresponds to the occurrence of the migration energies for certain values $(a - c)$. For simplicity, $c$ does not include a jumping atom. (For interpretation of the references to colour in this figure legend, the reader is referred to the web version of this article.)

**Fig. 11.** Histogram of normalised differences between the barriers of the same stable events calculated with and without tethering.

$(5, 1) \rightarrow (5, 4)$ process with a barrier $E_m(5, 1, 5, 4) = 0.43$ eV. If this process is excluded from the set, hence becoming effectively forbidden, the result shows a detachment of the tip from the substrate instead of its flattening. We simulated the tip geometry without a substrate to identify the events responsible for flattening. Only 0.008% of the events in the tip simulations led directly to flattening, the rest of the time the system spends on faceting, i.e. atoms moving along the crystallographic channels and hopping on facets.

We can also point out that more than 60% of the jumps are not advancing ones. The atoms are trapped between the same two configurations hopping back and forth. These trapping configurations usually have the same number of 1nn atoms. The barriers for those jumps are small, hence the probability for them to happen is high.

Even though the individual barriers might not correspond to reality in *Cu Set 1* and *Fe Set 1*, the overall trends of the barrier distributions predict well the kinetics of the event on Cu and Fe surfaces. Although, both sets include many forbidden barriers, the important events for nanocube formation and tip flattening simulations are included in the sets and have reasonable barriers, thus the overall evolution of the structures in the KMC simulations presented in [3,4] were reliable and compared well with the MD simulations and experiments. The fitted attempt frequencies for *Cu Set 1* and *Fe Set 1* also allowed us to obtain a time evolution comparable with the MD results in [3,4].

### 3.2. Tethering force approach for strong relaxation effects

We first analysed the effect of tethering on barriers of stable events. The histogram in Fig. 11 shows that the normalised differences between the barriers of most of the stable processes calculated with and without tethering are within 0.1 eV, which indicates that tethering does not affect the stable events significantly. However, there is around 4% of the processes with the normalised difference exceeding 0.5 eV.

We compared the barriers of the whole *Cu Set 1* and *Cu Set 2* in order to investigate the effect of tethering on the calculation of the barriers for the events involving also unstable configurations. The correlation of the barriers of both sets corresponding to the same $(a, b, c, d)$ combination is shown in Fig. 12. The largest disagreement between two sets are for the events with the very small barriers in *Cu Set 1*, i.e. spontaneous, shown in the insert of Fig. 12. Most of them correspond to $a \leqslant 3$. With the tethering force, the spontaneous events of *Cu Set 1* can be assigned finite barriers with the broad range of values up to 2 eV in *Cu Set 2*. Note, nevertheless, that even with the tethering force applied, the use of Eq. (2) was unavoidable for some barriers. However, the number of such barriers decreased to 283, with 211 barriers corresponding to the processes with $a \leqslant 3$, compared to 2676 spontaneous processes (out of which 2486 were for $a \leqslant 3$) in *Cu Set 1*. These 283 barriers are not included in the inserted picture in Fig. 12.

The tethering force approach allows us to calculate the barriers for all possible configurations in a rigid lattice. Thus, many $(a, b, c, d)$ jumps that were forbidden in *Cu Set 1*, are now allowed in *Cu Set*

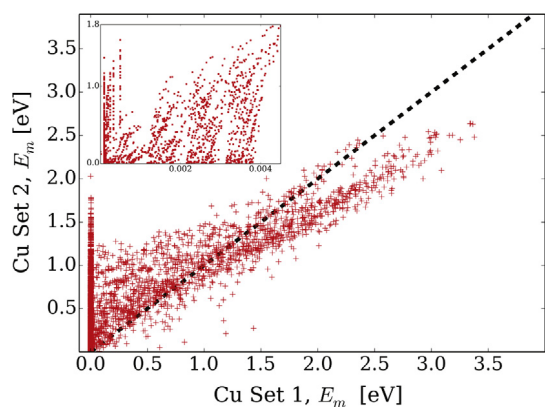

**Fig. 12.** Correlation of the barriers of *Cu Set 1* with the barriers of *Cu Set 2*. Inserted picture shows the barriers for the events with $E_m \leqslant 0.0047$ eV in *Cu Set 1*.

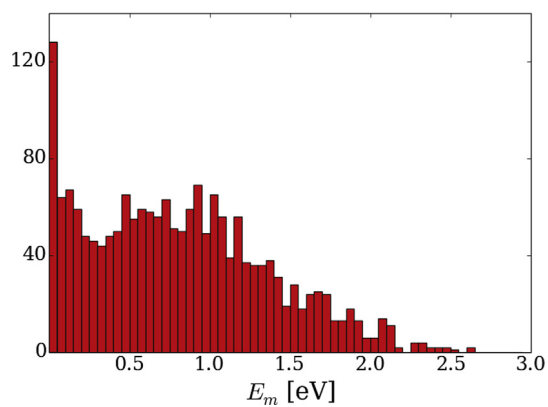

**Fig. 13.** Histogram of barriers of processes, which are unstable on a rigid lattice (forbidden in *Cu Set 1*), but calculated with the tethering force approach in *Cu Set 2*.

2. As it can be seen in Fig. 13, many of these unstable configurations are now allowed with high probabilities. One of the reasons why previously unstable configurations are now found to be stable and have low barriers is that the permutations in *Cu Set 2* were chosen based on the lowest potential energy of initial and final states and, thus, could differ from the permutations in *Cu Set 1*, which were selected at random. Another reason is that the tethering force approach is not capable to catch the processes with deep minima such as in Fig. 2d; instead a process like this is assigned with a small barrier. Whether these processes have impact on the KMC results or not depends on the considered system. We identified processes with such deep minima along the MEP between two rigid lattice positions in the post processing of the *Cu Set 2* and forbade them from happening in the tip flattening KMC simulations, which will be discussed below. Overall around 10% of the processes were forbidden in post processing, but it made little difference to the results of the KMC simulations of nanotips.

The spread of the barriers in *Cu Set 2* in Fig. 8c looks similar to the spread observed for *Cu Set 1* in Fig. 8b, but it has a much smoother transition from the region of events with $(a - c < 0)$ to $(a - c > 0)$. This reflects the consistency we had in choosing the permutations and treating the events involving unstable atoms in the calculations of the barriers of *Cu Set 2*.

We reproduced results of {110} tip flattening KMC simulations in [3] with *Cu Set 2* at 900 K. The profile of the tip flattening is similar: faceting of the tip near the foot occurred first and then adatoms from the top of the tip slid down towards the surface (see Fig. 14). The barrier for the same key events describing the sliding of adatoms downwards is $E_m(5, 1, 5, 4) = 0.47$ eV, which is only 0.04 eV higher than that in *Cu Set 1*. However, at temperatures above 900 K, the necking is emphasized and the nanotip detaches from the substrate instead of flattening down as was seen in simulations with *Cu Set 1*. The analysis of the occurred transitions showed that previously forbidden unstable processes in *Cu Set 1* (but now assigned with barriers using the tethering force approach in *Cu Set 2*) start to play a significant role in the system evolution at high temperatures. It should be noted that for such a Cu nanotip, with a thickness of 2 nm, the system will start to melt already at ~900 K due to finite size effects [44]. A rigid lattice model like Kimocs would not be able to accurately simulate molten system.

In order to estimate how the tethering force approach affects the time evolution, we have also compared the time scales of nanotip flattening obtained in KMC simulations with *Cu Set 1* and *Cu Set 2*. The simulations have shown a good agreement for the temperature 900 K: the flattening time of the 11 nm ⟨110⟩ nanotip with *Cu Set 1* was found to be $(2.0 \pm 0.20)$ μs and with *Cu Set 2* — $(5.8 \pm 0.44)$ μs.

For the tethering approach, we can conclude that it offers a solution for the parameterization of the events in a rigid lattice in KMC models, especially in those cases when it is not known *a priori* which jumps are crucial for surface evolution and it is important to generate as complete large tables of parameters as possible. The comparison between *Cu Set 1* and *Cu Set 2* indicates that the tethering force not only enables the calculation of the barriers in the unstable configurations (strong relaxation is expected in NEB), but also predicts well the relative distribution of the barriers, although some individual rigid lattice events can be overestimated due to their instability caused by relaxation effects. However, the barriers are not overestimated too much as the distribution of barriers is quite smooth.

### 3.3. Influence of the interatomic potential on the barrier calculations

Although we have performed most of our calculations of the barriers with the potentials described in Section 2.2, we analyse here the sensitivity of the barriers to the choice of the interatomic potential for the barrier calculations since this is essential for the validity of the KMC simulations.

The barrier calculation of 1024 possible permutations of hop-on jumps on Cu {100} surface, calculated with the CEM potential (see the next section), were repeated by constructing two new sets using the Mishin [45] and Sabochick-Lam [46] potentials. The average difference between the barriers calculated with CEM and the other two potentials is close to zero. Mishin potential barriers had a difference of $0.0002 \pm 0.09$ eV compared to CEM. The Sabochick-Lam potential resulted in a $0.008 \pm 0.11$ eV difference. Although the average difference is clearly not significant, we noticed that some of the rigid lattice adatom positions were identified as unstable during the NEB relaxation step with the Sabochick-Lam and Mishin potentials, whereas the same configurations were treated as stable with the CEM potential.

Here we note that KMC simulations rely mostly on the relative distribution of barriers of different events rather than on the absolute values of the barriers. Fig. 15 shows that the trend in the barriers depending on the change in the number of 1nn atoms of the jumping atom before and after the jump is quite similar for the Mishin and Sabochick-Lam potentials. At the same time, the CEM potential results in a smoother relative distribution of the barriers.

Based on our analysis, we conclude that although the relative distribution of the migration barriers calculated by using different interatomic potentials can be similar, still some differences may exist. This is why it is important to verify the values of crucial barriers for the simulated processes in order to ensure that they are predicted sufficiently accurately by the potential in use. For instance, if adatom diffusion is studied, the potential which was used to calculate the energy barriers should be able to predict the barriers for all 1024 hop-on jumps for different arrangements of neighbouring atoms. In case the event cannot be calculated, the use of a tethering force is recommended as it will keep the unstable atoms in their positions, allowing for some relaxation effects during the transition. This way, the relative distribution of the barriers can still be obtained in a reliable manner, which will enable the accurate prediction of the overall evolution of the system.

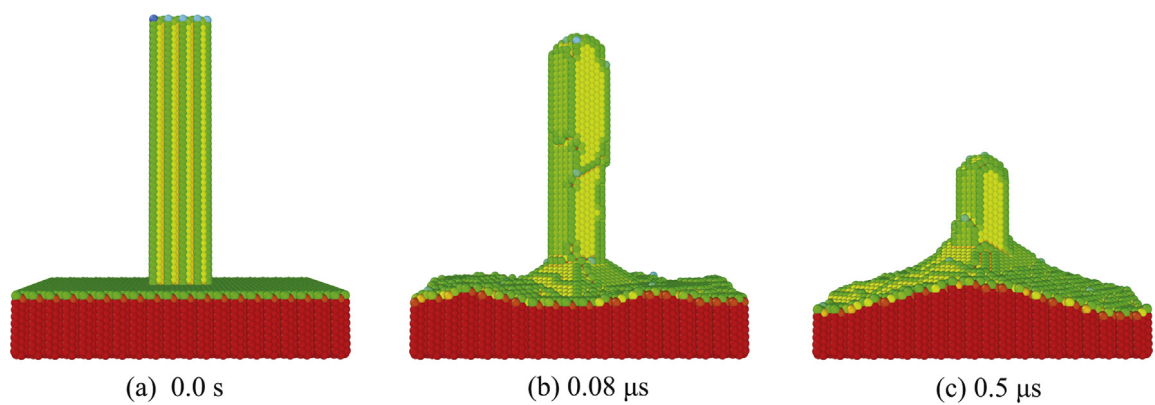

(a) 0.0 s          (b) 0.08 μs          (c) 0.5 μs

**Fig. 14.** The profile and corresponding time of the 11 nm height and 2 nm wide ⟨110⟩ nanotip, which has flattened in the Kimocs simulations at 900 K with *Cu Set 2*. Colours correspond to coordination numbers. (For interpretation of the references to colour in this figure legend, the reader is referred to the web version of this article.)

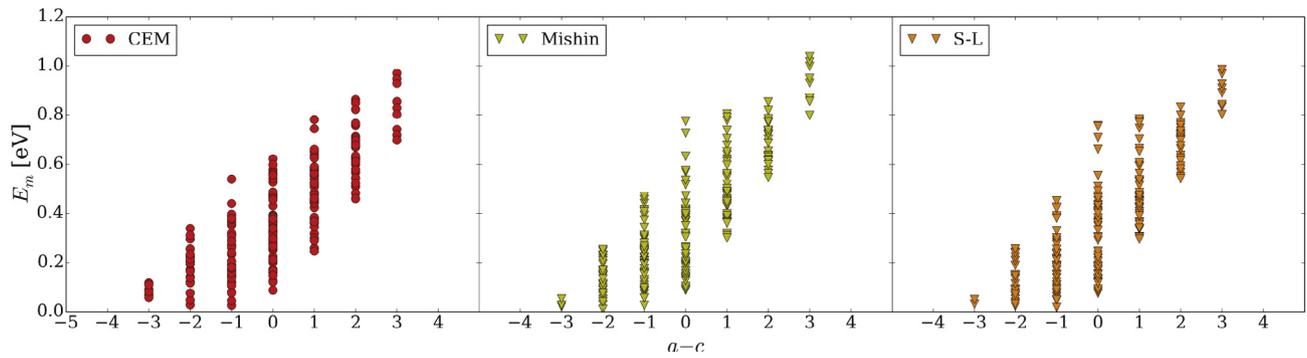

**Fig. 15.** Migration energy barriers vs. the change in the number of 1nn of the initial and final positions of the adatom hopping-on on Cu {100} surface calculated with different interatomic potentials. Left - Corrected Effective Medium Theory (CEM) potential [23], middle - Mishin [45], right - Sabochick-Lam [46].

### 3.4. 4D approach versus a 26D description

In the 4D approach, as it has been discussed before, events of diffusion jumps are described only by counting the numbers of 1nn and 2nn atoms around the initial and final sites of the jumping atom and the actual position of the neighbours are thus ignored in the characterization. Subsequently, an uncertainty of the value of the calculated barrier is expected. In some configurations, the number of permutations corresponding to the same combination of 1nn and 2nn neighbours can be significant. As it was mentioned above, we dealt with this problem by following two simple principles. A permutation representing a family of events described by the same $(a, b, c, d)$ numbers, was chosen either (i) randomly, as in *Cu Set 1* and *Fe Set 1*, or (ii) according to the lowest energy state of the initial and final configurations, as in *Cu Set 2*. As an example, Fig. 16 shows minimum energy paths for three different permuta-tions of the same $(5, 3, 7, 3)$ hop-on jump on Cu {100} surface. As it can be seen, one of the permutations shows significantly different MEP with the lowest energy barrier being 0.07 eV, while the other two, — the black and red curves — result in more similar barriers of 0.28 eV and 0.26 eV, respectively. The $(5, 3, 7, 3)$ configuration has overall 25 atomic permutations if only considering 1nn and 2nn atoms in the same layer of the initial and final positions of the hopping atom, while all the neighbouring atoms in the layer below are present and thus do not affect the permutations). The average barrier for the $(5, 3, 7, 3)$ configuration obtained over all 25 permutations is $0.23 \pm 0.1$ eV. The permutation corresponding to the lowest energy initial and final states is assigned with 0.27 eV by the tethering force approach in *Cu Set 2*. However, the randomly chosen permutation happened to be described by the barrier 0.07 eV, which was used in *Cu Set 1*. We chose the most dramatic case for illustration of this uncertainty, which actually indicates that in the approach chosen initially, i.e. adopting a single value of the barrier for all $(a, b, c, d)$ events, which was calculated for a randomly selected permutation, may affect the simulation results. If a key process is assigned a barrier with too imprecise value, artefacts may appear in the simulations. The validation of the parameterization — and thereby making sure it does not produce artefacts — can usually be done by comparing the KMC results for small systems with MD, as it was done in [3].

In our implementation, surfaces with different crystallographic orientations can be simulated, since all of them can be described within the same 4D formalism. Even if the positions of neighbouring atoms on a rigid lattice are not the same in different facets, the barriers fit together remarkably well.

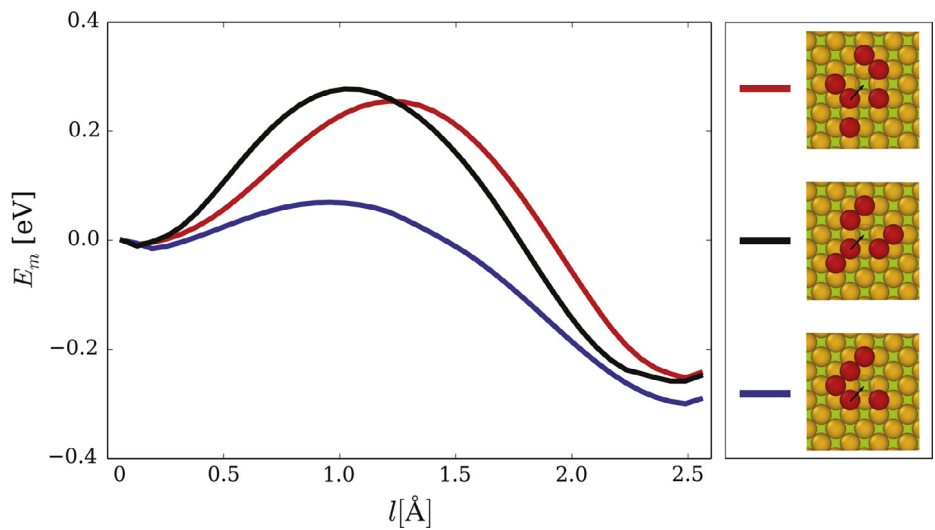

**Fig. 16.** Minimum energy paths of the different permutations of the $(5, 3, 7, 3)$ jump. $l$ is the distance from the initial position of the jumping atom.

In the following, we will give a quantitative estimate of the uncertainty, which arises when different permutations for the same number of neighbours are ignored. Consider the case of hop-on jumps on a {100} FCC surface, where the layer below the jumping atom is filled completely in all considered permutations. In this case, the only neighbour sites that can change are the ten 1nn and 2nn atom positions around the initial and final positions of the jumping atom. The total number of different permutations in this case is $2^{10} = 1024$, based on whether these 10 sites are occupied or vacant. In the Kimocs description, there are overall 196 $(a, b, c, d)$ events with 1024 permutations on the {100} FCC surface. The energy barriers, which were calculated for all permutations in these 196 configurations for Cu are shown in Fig. 17. Most of the barriers are within the range of [0.1, 0.6] eV. The histogram also includes the barriers for an atom diffusion on the close-packed {111} Cu surface. There are only 23 combinations of $(a, b, c, d)$ events on the {111} surface with most of the barriers being less than 0.1 eV. Note that our model does not include the migration to HCP sites on the {111} surface. The average standard deviation of the energy barriers for the 196 configurations of the {100} surface was found to be 0.13 eV or 14.8%, which can be used as the estimate of the accuracy of the energy barrier parameterization and, consequently, of our KMC model in general.

To further analyse the effect of permutations on the value of the barrier, we compared the barriers in *Cu Set 2*, obtained in the 4D parameterization scheme, and *Cu Set 3*, obtained in the full 26D parameterization scheme. These are comparable as the barriers in both sets were obtained with the application of the tethering force. The results of this comparison are presented in Fig. 18. Here we plot the values of the barriers corresponding to the same $(a, b, c, d)$ events as well, all of them will lie along the dashed line. In *Cu Set 3*, many barriers correspond to the same $(a, b, c, d)$ description due to different permutations. To enable the comparison, we averaged all the barriers, which belong to the same family of $(a, b, c, d)$ and plotted the averaged value against the value found in *Cu Set 2* for the same $(a, b, c, d)$ event. The standard deviation of the mean value shown around the point to indicate how broadly the values of the barriers fluctuate with different permutations. As one can see, the strong fluctuations of the barriers around the mean values are mostly observed for some individual events, while for the vast majority of events, these fluctuations are not strong and overall all the dots cluster quite closely to the dashed line. This confirms that the 4D parameterization scheme is able to capture the relative distribution of barriers quite well.

Our KMC simulations of flattening of Cu nanotips using *Cu Set 1* and *Cu Set 2* showed that 4D parameterization of atomic jumps was sufficient to obtain good agreement with MD simulations. In Section 3.1.2, we showed that an important role is played by the surface events in simulations of nanotips, i.e. the events that have less than ten 1nn and 2nn atoms. As we showed here, ignoring possible permutations did not introduce a significant uncertainty in barriers of surface events due to the small number of permutations available for such events. We note here that ignoring permutations in *Fe Set 1* for the Fe self-diffusion on flat surfaces did not affect the results obtained for growing iron nanocubes at low temperatures either, since the surface events played a crucial role also in these simulations. On the other hand, as we have shown in Fig. 18, in some cases the 4D description of the events may fail to capture the physics properly as the permutations hidden within such an approach may play a crucial role in some specific cases. For example, it was not possible to include the exchange events on Fe {100} surface without introducing artefacts on {110}, thus, at high temperatures, we obtained quite different results in KMC simulations compared to MD simulations.

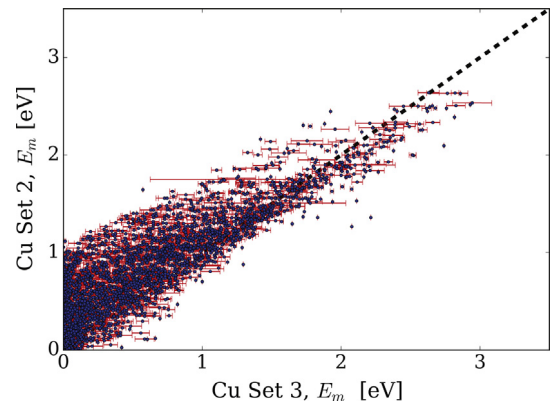

**Fig. 18.** Cu barriers of *Cu Set 3* in comparison with *Cu Set 2*.

## 4. Conclusions

The rigid lattice approximation allows to create fast and efficient Kinetic Monte Carlo models that are capable to simulate big systems of millions of atoms and reach much longer time scales than with e.g. molecular dynamics. The crucial part for any atomistic Kinetic Monte Carlo model is the set of thousands or even millions of transition barriers that defines the probabilities for any atom to make a jump in the system. These barriers need to be calculated separately — in our approach using molecular dynamics in combination with the Nudged Elastic Band method — and then parameterized and tabulated for use in the Kinetic Monte Carlo simulations.

In this paper, we have discussed different approaches for calculating and compiling such barrier sets for FCC and BCC metals. We have shown that many of the problems with a rigid lattice parameterization can be overcome by applying a tethering force in the Nudged Elastic Band calculations of the barriers; these additional spring forces prevent the atoms to relax to a position too far from the lattice positions during the relaxation and the Nudged Elastic Band calculation of the barrier.

We have calculated three sets of barriers for copper using three different approaches, including the tethering force approach, and also sets for iron. Sets of barriers can be found in [29,30]. We have analysed the distributions of the barriers in each set as well as the importance of individual barriers in Kinetic Monte Carlo simulations of surface processes. Our barrier sets of thousands of barriers for Cu and Fe could in principle be used by any Kinetic Monte Carlo model for atom migration processes in a rigid lattice.

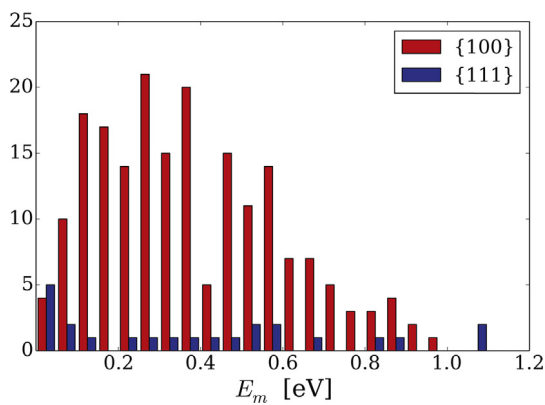

**Fig. 17.** The occurrence of the migration energies of adatom diffusion on {100} and {111} Cu surfaces.

We have shown that tethering force approach in the calculation of the migration barriers does not significantly change the physical outcome of Kinetic Monte Carlo simulations where sets of thousands of such barriers are used. On the contrary, this new approach provides a systematic way to calculate all barriers needed for Kinetic Monte Carlo simulations of a single-metal system; especially in systems where surface processes are important.

## Acknowledgement

The authors would like to thank Dr. Antti Kuronen for providing his implementation of the NEB method for PARCAS. This work was supported by CERN under the K-contract KE2488/BE/CLIC. E. Baibuz was supported by a CERN K-contract and the doctoral program MATRENA of the University of Helsinki. J. Lahtinen was supported by a CERN K-contract. V. Jansson was supported by Academy of Finland (Grant No. 285382). F. Djurabekova acknowledges gratefully the financial support of Academy of Finland (Grant No. 269696) and MEPhI Academic Excellence Project – Russia (Contract No. 02.a03.21.0005). V. Zadin and S. Vigonski were supported by Estonian Research Council grants PUT 57 and PUT 1372. Computing resources were provided by the Finnish IT Center for Science (CSC) and the Finnish Grid and Cloud Infrastructure (persistent identifier urn:nbn:fi:research-infras-2016072533).